# THE ROLE OF TIME
# IN
# PHYSICAL COSMOLOGY

**D.S. SALOPEK**

*Department of Physics, University of Alberta, Edmonton, Canada T6G 2J1*

**Abstract**

Recent advances in observational cosmology are changing the way we view the nature of time. In general relativity, the freedom in choosing a time hypersurface has hampered the implementation of the theory. Fortunately, Hamilton-Jacobi theory enables one to describe all time hypersurfaces on an equal footing. Using an expansion in powers of the spatial curvature, one may solve for the wavefunctional in a semiclassical approximation. In this way, one may readily compare predictions of various inflation models with observations of microwave background anisotropies and galaxy clustering.

## 1 Introduction

In the standard cosmological formulation of Einstein's theory of gravity, one ordinarily solves the field equations for the 4-metric $g_{\mu\nu}$. In a more elegant although equivalent approach, it has been recommended [1]-[5] that one solve the Hamilton-Jacobi (HJ) equation for general relativity which governs the evolution of the generating functional $\mathcal{S}$. By adopting this method, many advantages have be gained:

**(1) Avoiding Gauge Problems.** Neither the lapse $N$ nor the shift $N_i$ appear in the HJ equation. The generating functional depends only on the 3-metric $\gamma_{ij}$ (as well any matter fields that may be present). Hence, the structure of the HJ equation is conceptually simpler than that of the Einstein field equations. For instance, one is able to avoid the embarrassing problem of picking a gauge in general relativity. As a corollary, one obtains a deep appreciation for the nature of cosmic time.

**(2) Solution of Constraint Equations.** One may solve the constraint equations of general relativity in a systematic manner, even in a nonlinear setting [4], [6]. The momentum constraint is easy to solve using HJ theory: one simply constructs $\mathcal{S}$ using integrals of the 3-curvature over the entire 3-geometry. The energy constraint may solved in solving by expanding in powers of the 3-curvature ('spatial gradient expansion').

**(3) Primitive Quantum Theory of Gravity.** Solutions of the HJ equation may be interpreted as the lowest order contribution to the wavefunctional for an expansion in powers of $\hbar$. One may describe some quantum processes such as the initial 'ground state' of the Universe or tunnelling through a potential barrier.

Hence, HJ theory has proven to be a particularly powerful tool for the cosmologist. In this article, I will discuss our current understanding of the theory. In addition, I review the observational status of three models of cosmological inflation:

(1) inflation with an exponential potential ('power-law inflation') [7], [8] which arises naturally from Induced Gravity [9] or Extended Inflation [10];

(2) inflation with a cosine potential ('natural inflation') [11] where the inflaton is a pseudo-Goldstone boson;

(3) inflation with two scalar fields ('double inflation') [12], [9] where there are two periods of inflation.

## 2 Hamilton-Jacobi Equation for General Relativity

The Hamilton-Jacobi equation for general relativity is derived using a Hamiltonian formulation of gravity. One first writes the line element using the ADM 3+1 split,

$$ds^2 = \left(-N^2 + \gamma^{ij} N_i N_j\right) dt^2 + 2N_i dt\, dx^i + \gamma_{ij} \; dx^i dx^j \; , \tag{1}$$

where $N$ and $N_i$ are the lapse and shift functions, respectively, and $\gamma_{ij}$ is the 3-metric. Hilbert's action for gravity interacting with a scalar field becomes

$$\mathcal{I} = \int d^4x \left(\pi^\phi \dot{\phi} + \pi^{ij} \dot{\gamma}_{ij} - N\mathcal{H} - N^i \mathcal{H}_i\right) . \tag{2}$$

The lapse and shift functions are Lagrange multipliers that imply the energy constraint $\mathcal{H}(x) = 0$ and the momentum constraint $\mathcal{H}_i(x) = 0$.

The object of chief importance is the generating functional $\mathcal{S} \equiv \mathcal{S}[\gamma_{ij}(x), \phi(x)]$. For each scalar field configuration $\phi(x)$ on a space-like hypersurface with 3-geometry described by the 3-metric $\gamma_{ij}(x)$, the generating functional associates a complex number. The generating functional is the 'phase' of the wavefunctional in the semi-classical approximation: $\Psi \sim e^{i\mathcal{S}}$. (The prefactor is neglected here although it has important implications for quantum cosmology [13].) The probability functional, $\mathcal{P} \equiv |\Psi|^2$, is given by the square of the wavefunctional.

Replacing the conjugate momenta by functional derivatives of $\mathcal{S}$ with respect to the fields,

$$\pi^{ij}(x) = \frac{\delta \mathcal{S}}{\delta \gamma_{ij}(x)} , \qquad \pi^{\phi}(x) = \frac{\delta \mathcal{S}}{\delta \phi(x)} , \tag{3}$$

and substituting into the energy constraint, one obtains the Hamilton-Jacobi equation,

$$\begin{aligned} \mathcal{H}(x) = \quad & \gamma^{-1/2} \frac{\delta \mathcal{S}}{\delta \gamma_{ij}(x)} \frac{\delta \mathcal{S}}{\delta \gamma_{kl}(x)} \left[ 2\gamma_{il}(x)\gamma_{jk}(x) - \gamma_{ij}(x)\gamma_{kl}(x) \right] \\ & + \frac{1}{2}\gamma^{-1/2} \left( \frac{\delta \mathcal{S}}{\delta \phi(x)} \right)^2 + \gamma^{1/2} V(\phi(x)) \\ & - \frac{1}{2}\gamma^{1/2} R + \frac{1}{2}\gamma^{1/2}\gamma^{ij}\phi_{,i}\phi_{,j} = 0 , \end{aligned} \tag{4}$$

which describes how $\mathcal{S}$ evolves in superspace. $R$ is the Ricci scalar associated with the 3-metric, and $V(\phi)$ is the scalar field potential. In addition, one must also satisfy the momentum constraint

$$\mathcal{H}_i(x) = -2 \left( \gamma_{ik} \frac{\delta \mathcal{S}}{\delta \gamma_{kj}(x)} \right)_{,j} + \frac{\delta \mathcal{S}}{\delta \gamma_{lk}(x)} \gamma_{lk,i} + \frac{\delta \mathcal{S}}{\delta \phi(x)} \phi_{,i} = 0 , \tag{5}$$

which legislates gauge invariance: $\mathcal{S}$ is invariant under reparametrizations of the spatial coordinates. (Units are chosen so that $c = 8\pi G = \hbar = 1$). Since neither the lapse function nor the shift function appears in eqs.(4,5) the temporal and spatial coordinates are *arbitrary*: HJ theory is *covariant*.

## 3 Spatial Gradient Expansion

As a first step in solving eqs.(4,5), I will expand the generating functional

$$\mathcal{S} = \mathcal{S}^{(0)} + \mathcal{S}^{(2)} + \mathcal{S}^{(4)} + \ldots , \tag{6}$$

in a series of terms according to the number of spatial gradients that they contain. The invariance of the generating functional under spatial coordinate transformations suggests a solution of the form,

$$\mathcal{S}^{(0)}[\gamma_{ij}(x), \phi(x)] = -2 \int d^3x \, \gamma^{1/2} H \left[ \phi(x) \right] , \tag{7}$$

for the zeroth order term $\mathcal{S}^{(0)}$. The function $H \equiv H(\phi)$ satisfies the separated HJ equation of order zero [5],

$$H^2 = \frac{2}{3} \left( \frac{\partial H}{\partial \phi} \right)^2 + \frac{1}{3} V(\phi) , \tag{8}$$

which is an ordinary differential equation. Note that $\mathcal{S}^{(0)}$ contains no spatial gradients.

In order to compute the higher order terms, one introduces a change of variables, $(\gamma_{ij}, \phi) \rightarrow (f_{ij}, u)$:

$$u = \int \frac{d\phi}{-2\frac{\partial H}{\partial \phi}} , \quad f_{ij} = \Omega^{-2}(u)\, \gamma_{ij} , \tag{9}$$

where the conformal factor $\Omega \equiv \Omega(u)$ is defined through

$$\frac{d \ln \Omega}{du} \equiv -2 \frac{\partial H}{\partial \phi} \frac{\partial \ln \Omega}{\partial \phi} = H . \tag{10}$$

in which case the equation for $\mathcal{S}^{(2m)}$ becomes

$$\left. \frac{\delta \mathcal{S}^{(2m)}}{\delta u(x)} \right|_{f_{ij}} + \mathcal{R}^{(2m)}[f_{ij}(x), u(x)] = 0 . \tag{11}$$

The remainder term $\mathcal{R}^{(2m)}$ depends on some quadratic combination of the previous order terms (*i.e.*, it may be written explicitly [3]). For example, for $m = 1$, it is

$$\mathcal{R}^{(2)} = \frac{1}{2} \gamma^{1/2} \gamma^{ij} \phi_{,i} \phi_{,j} - \frac{1}{2} \gamma^{1/2} R . \tag{12}$$

Eq.(11) has the form of an infinite dimensional gradient. It may be integrated directly using a line integral:

$$\mathcal{S}^{(2m)} = - \int d^3x \int_0^1 ds\ u(x)\ \mathcal{R}^{(2m)}[f_{ij}(x), s u(x)] . \tag{13}$$

For simplicity, the contour of integration was chosen to be a straight line in superspace. As long as the end points are fixed, the line integral is independent of the contour choice which corresponds to the choice of time hypersurface. This property goes a long way in illuminating the nature of time in general relativity.

Typically, $\mathcal{S}^{(2m)}$ is an integral of terms which contain the Ricci tensor and derivatives of the scalar field [3]. For $m = 1$, one determines that

$$\mathcal{S}^{(2)}[f_{ij}(x), u(x)] = \int d^3x f^{1/2} \left[ j(u) \widetilde{R} + k(u) f^{ij} u_{,i} u_{,j} \right] . \tag{14}$$

$\widetilde{R}$ is the Ricci scalar of conformal 3-metric $f_{ij}$ appearing in eq.(9). The $u$-dependent coefficients $j$ and $k$ are,

$$j(u) = \int_0^u \frac{\Omega(u')}{2}\, du' + F, \qquad k(u) = H(u)\, \Omega(u) , \tag{15}$$

where $F$ is an arbitrary constant.

### 3.1 Characteristics of Cosmic Time

The generalization of the spatial gradient expansion to multiple scalar fields is non-trivial [2]. In this case, one employs the method of characteristics for solving the linear partial differential equation that appears. For a single scalar field $\phi$ in a HJ description, it is obvious to use some function of $\phi$ as the integration parameter. In order to facilitate the integration of $\mathcal{S}^{(2)}$ for multiple fields, I recommend using the scale factor, $\Omega \equiv \Omega(\phi_a)$, which is a specific function of the scalar fields. A rigorous proof of the the integrability condition for the spatial gradient expansion is given in ref.[2].

# 4 Quadratic Curvature Expansion

In order to describe the fluctuations arising during the inflationary epoch, it is necessary to sum an infinite subset [1] of the terms $\mathcal{S}^{(2m)}$. In this case, one makes an Ansatz which includes all terms which are quadratic in the Ricci tensor $\tilde{R}_{ij}$ of the conformal 3-metric $f_{ij}(x)$:

$$\mathcal{S} = \mathcal{S}^{(0)} + \mathcal{S}^{(2)} + \mathcal{Q}\,; \tag{16}$$

here the quadratic functional $\mathcal{Q}$ is

$$\mathcal{Q} = \int d^3x f^{1/2} \left[ \tilde{R}\ \ \widehat{S}(u, \tilde{D}^2)\ \tilde{R} + \tilde{R}^{ij}\ \widehat{T}(u, \tilde{D}^2)\ \tilde{R}_{ij} - \frac{3}{8} \tilde{R}\ \widehat{T}(u, \tilde{D}^2)\ \tilde{R} \right], \tag{17}$$

where $\widehat{S}(u, \tilde{D}^2)$ and $\widehat{T}(u, \tilde{D}^2)$ are differential operators which are also functions of $u$. $\widehat{S}$ and $\widehat{T}$ describe scalar and tensor fluctuations, respectively. $\tilde{D}^2$ is the Laplacian operator with respect to the conformal 3-metric.

## 4.1 Multiple Fields

Once again, the case for two fields [2] is more complicated (the extension to any additional fields is straightforward). After choosing $\Omega(x)$ as the integration variable, one computes a constant of integration $e(x)$ in the long-wavelength approximation. One replaces $\widehat{S}$ by a matrix operator $\widehat{S}_{ab}$, $a, b = 1, 2$, which is a function of $\Omega(x)$ and $e(x)$. The scalar operator $\widehat{S}_{ab}$ is then sandwiched between the vector $[\tilde{R}, \tilde{D}^2 e]$ and its transpose in eq.(17).

# 5 Comparison with Large-Angle Microwave Background Fluctuations and Galaxy Correlations

Using HJ theory, I will compare the cosmological implications of three inflationary models: Model 1 — power-law inflation; Model 2 — natural inflation, and Model 3 — double inflation. All models will be normalized using the latest data for large angle microwave background anisotropies determined by COBE [14]: $\sigma_{sky}(10^0) = 30.5 \pm 2.7\mu K$ (68% confidence level).

It is conventional to parametrize the primordial scalar fluctuations arising from inflation by $\zeta$ which is proportional to the metric perturbation on a comoving time hypersurface [5]. The power spectra for $\zeta$ are shown in Figs.(1a), (2a), (3a), for Models 1, 2, 3, respectively. Both power-law inflation and natural inflation yield power spectra $\mathcal{P}_\zeta(k)$ for $\zeta$ that are pure power-laws:

$$\mathcal{P}_\zeta(k) = \mathcal{P}_\zeta(k_0)\left(\frac{k}{k_0}\right)^{n-1} . \tag{18}$$

The spectral index for scalar perturbations is denoted by $n$, and $n = 1$ describes the flat Zel'dovich spectrum. The simplest models arising from inflation are characterized by $n < 1$. The normalization of the spectra differs for the first two models since gravitational waves may contribute significantly to COBE's signal for the power-law inflation model [8]. The primordial power spectrum for double inflation is not a pure power-law. (In Fig.(3a), the primordial fluctuations for inflation with a single field having a quadratic potential is also shown.)

The power spectra $\mathcal{P}_\delta(k)$ for the linear density perturbation $\delta = \delta\rho/\rho$ at the present epoch are shown in Figs.(1b), (2b), (3b). The data points with error bars are determined from the clustering of galaxies [15]. I have assumed that the evolution of the fluctuations is described by the cold-dark-matter transfer function [16] where the present Hubble parameter is taken to be $H_0 = 50$ km s$^{-1}$Mpc$^{-1}$.

For power-law inflation, the best fit is given by $n = 0.9$ (bold curve in Fig.(1b)). However, $n = 0.8$ gives the best fit for natural inflation. From the galaxy data, there is not much difference between the best fits of these two models. One hopes to discriminate these models further when intermediate angle microwave background fluctuations [17] become more precise. Model 3, double inflation model is not particularly attractive since it requires three parameters whereas the previous two models each required one less. For the choice of double inflation parameters advocated by Peter *et al* [18], there is not much advantage over the simpler models of power-law inflation and double inflation [2] (see Fig.(3b)).

## 6 Summary

The question of time choice in general relativity is a difficult one, particularly for the quantum theory [19]. For semi-classical problems of interest to observational cosmology, one may construct a covariant formalism which treats all time choices on an equal footing. Although not quite perfect, reasonable fits of microwave background anisotropies and galaxy clustering may be obtained by power-law inflation with a spectral index of $n = 0.9$, or by natural inflation with a spectral index $n = 0.8$. If wishes to pay the price for an additional parameter, double inflation is also adequate.

# 7 Acknowledgments

I thank J.M. Stewart for a fruitful collaboration on some of the topics discussed. I also thank Don Page for useful discussions. This work was supported by NSERC and CITA of Canada.

# 8 Figure Captions

**Figs. (1a), (2a), (3a)**: Shown are the power spectra $\mathcal{P}_\zeta(k)$ for zeta, which describes the primordial scalar perturbations arising from inflation. Three plausible models of inflation are considered: (1) power-law inflation, (2) natural inflation and (3) double inflation. The first two models require two parameters: an arbitrary normalization factor and a spectral index $n$, where $n-1$ is the slope of the spectrum in a log-log plot. Double inflation is a three parameter model. For each model, the normalization is fixed by large angle microwave background anisotropies.

**Figs. (1b), (2b), (3b)**: For the present epoch, the power spectra for the linear density perturbation $\delta\rho/\rho$ are shown for the same models of Fig. (a). The data points with error bars are the observed power spectrum derived from galaxy surveys. For power-law inflation, the best fit (bold curve) is obtained with a spectral index of $n = 0.9$, whereas $n = 0.8$ yields the best fit for natural inflation. (Gravitational waves are important for power-law inflation but not for natural inflation.) Using an additional parameter, double inflation also gives a reasonable fit.

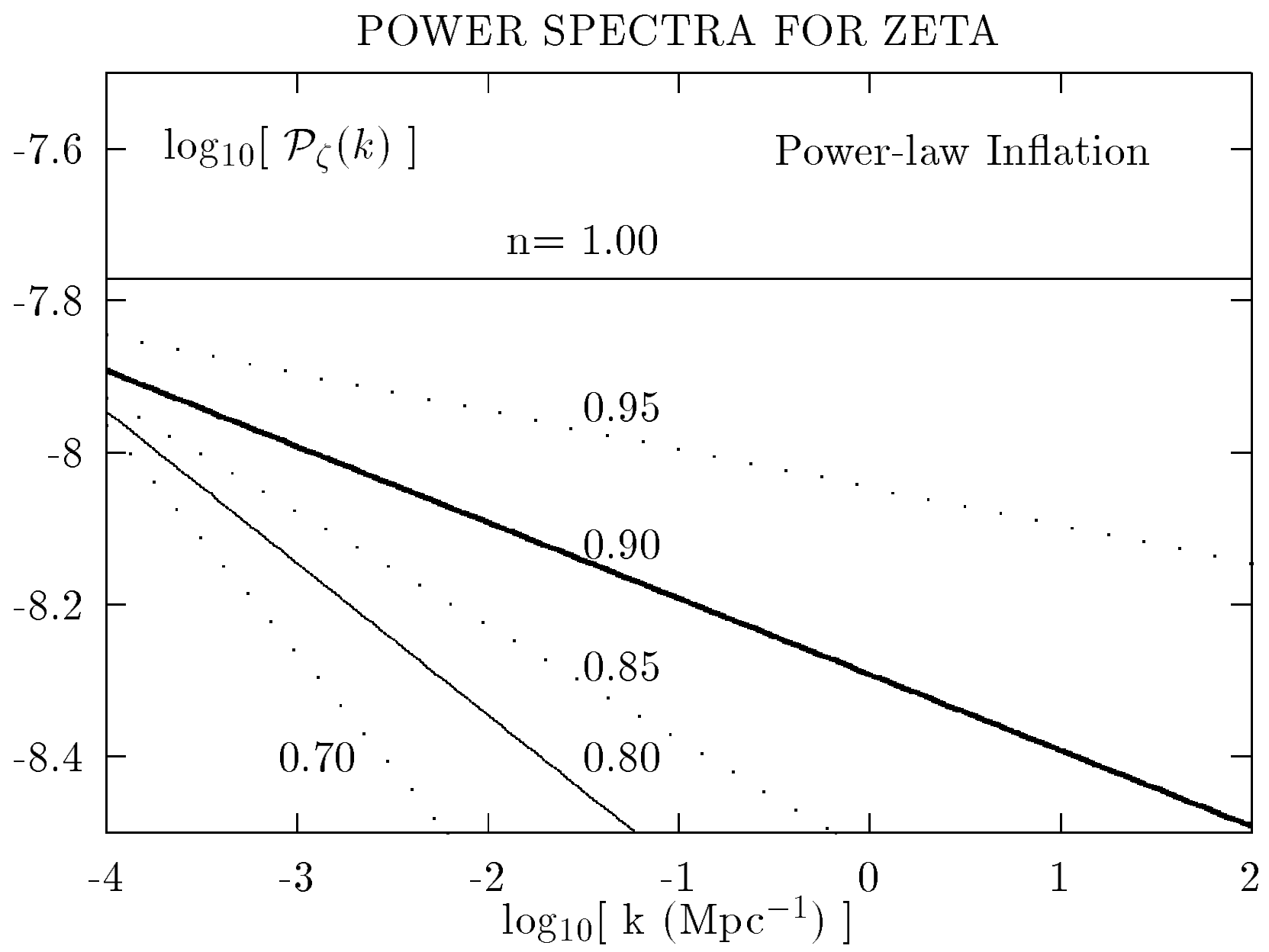

**Fig.(1a)**

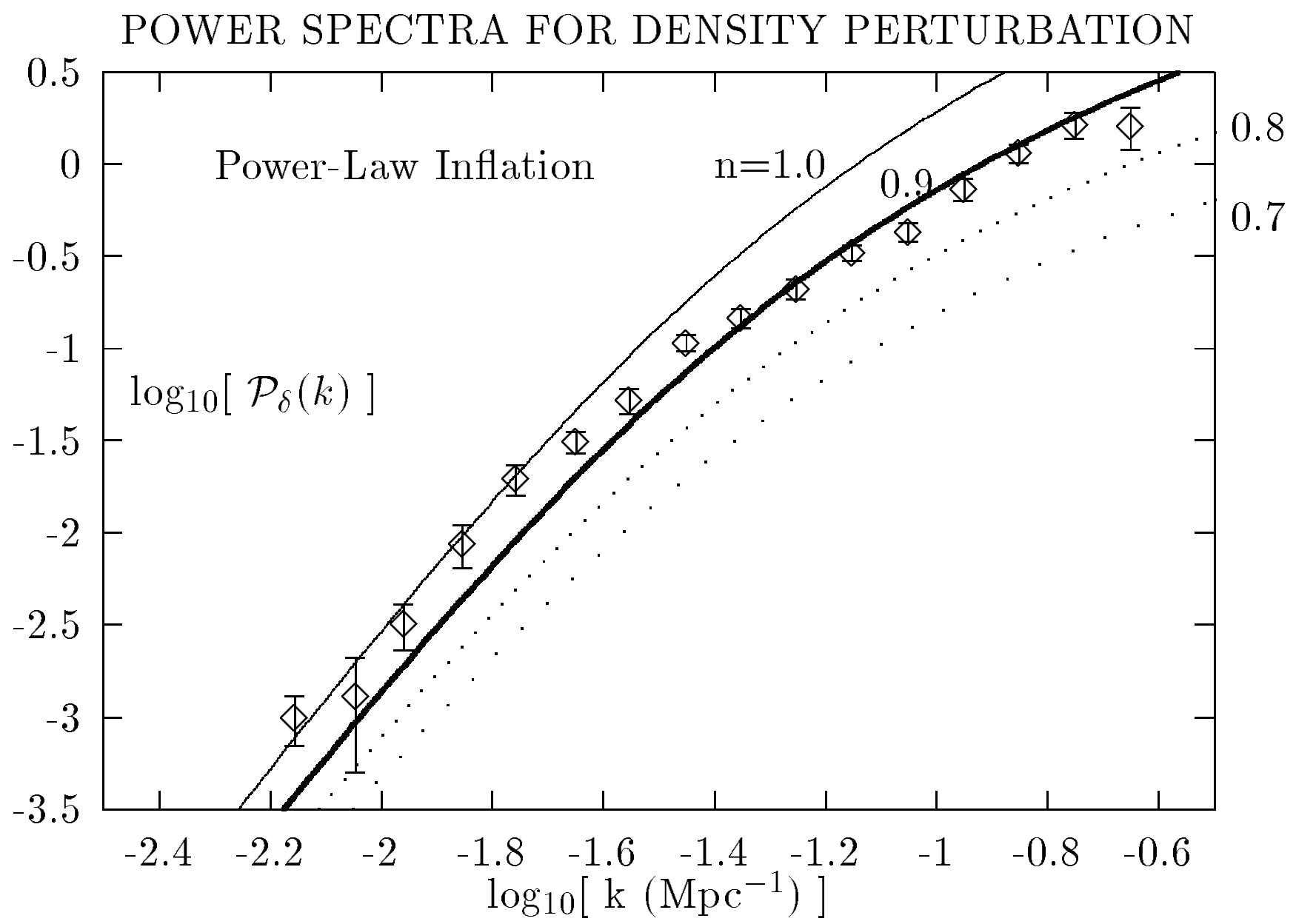

**Fig.(1b)**

Fig.(2a)

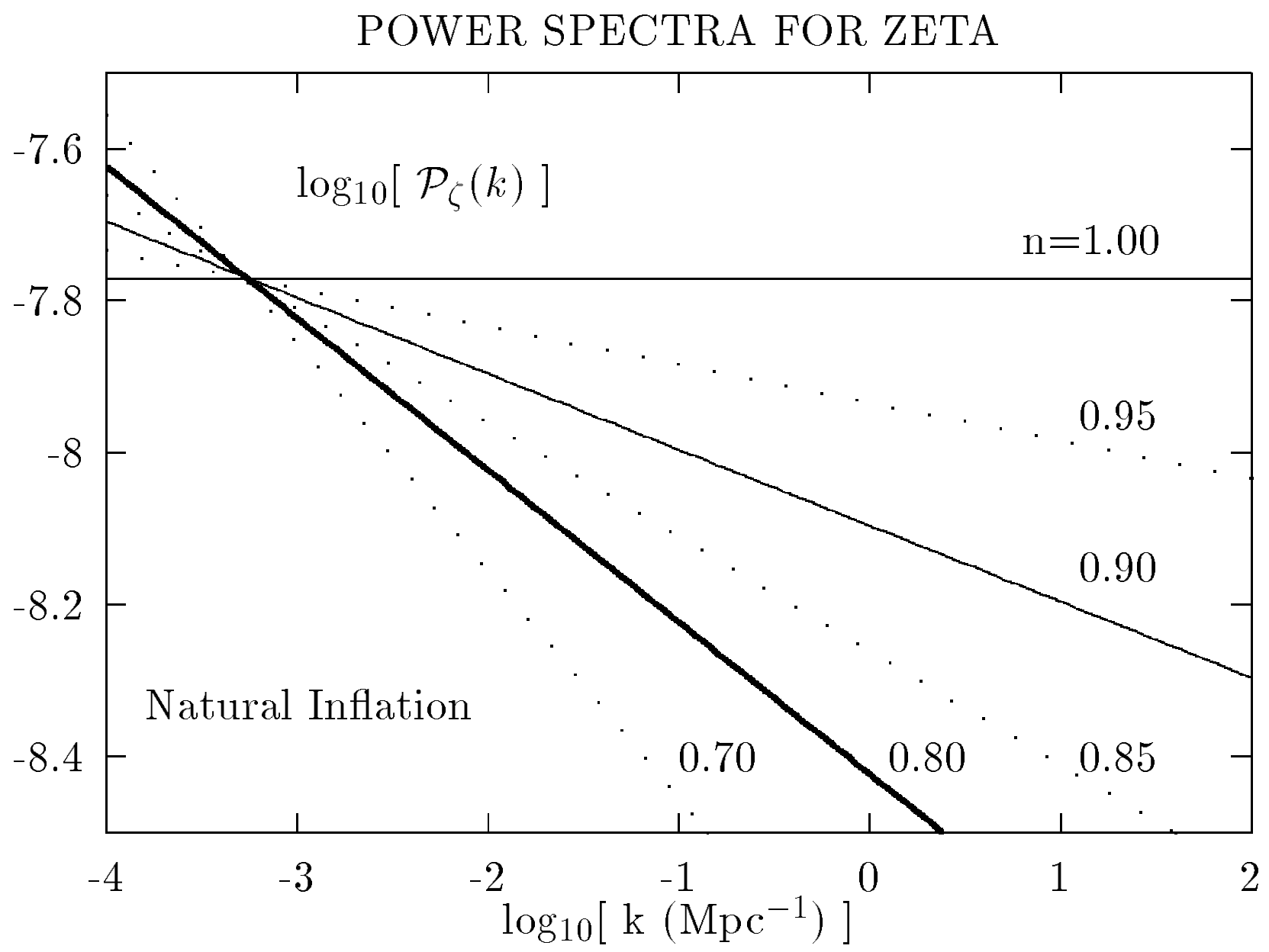

Fig.(2b)

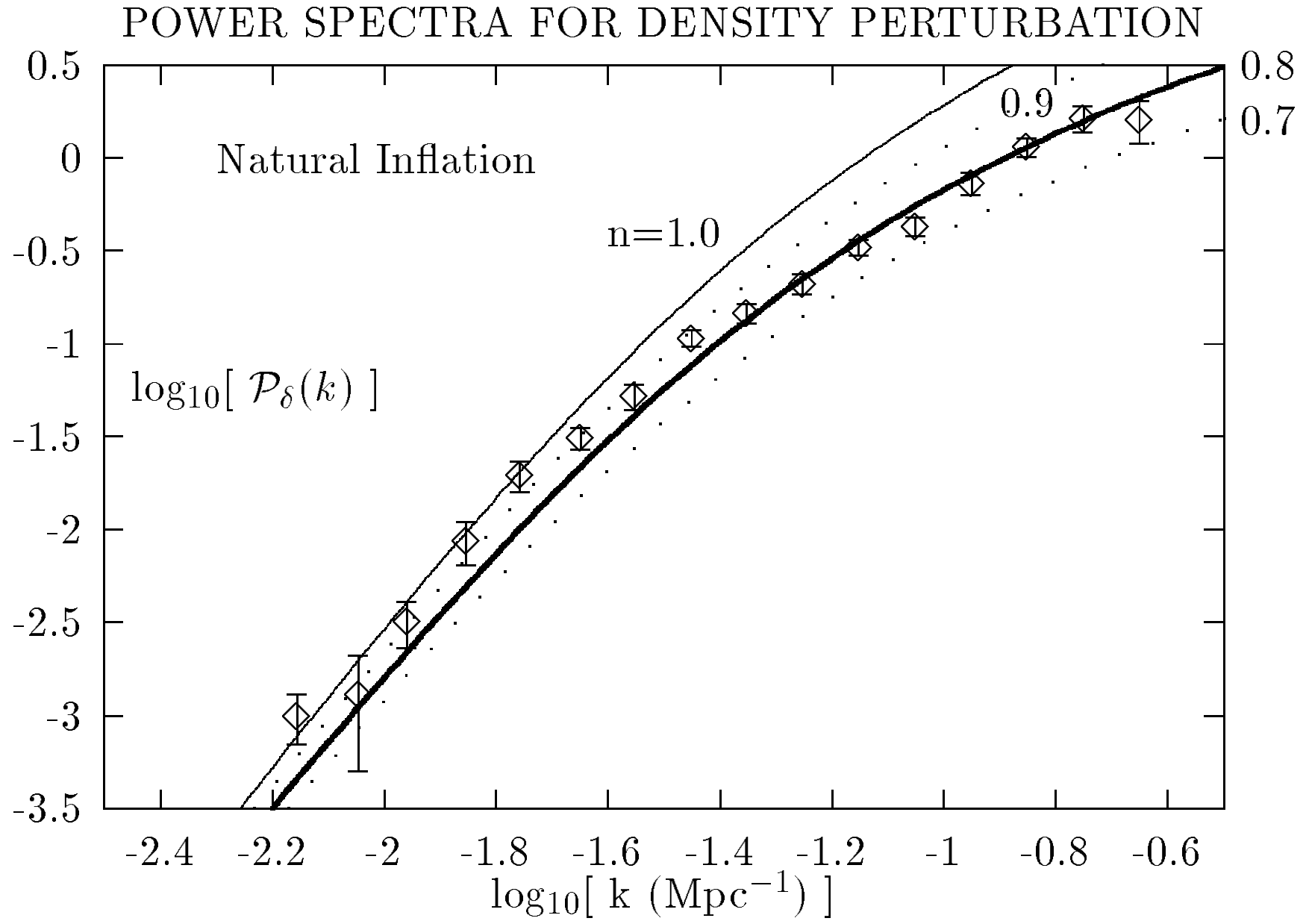

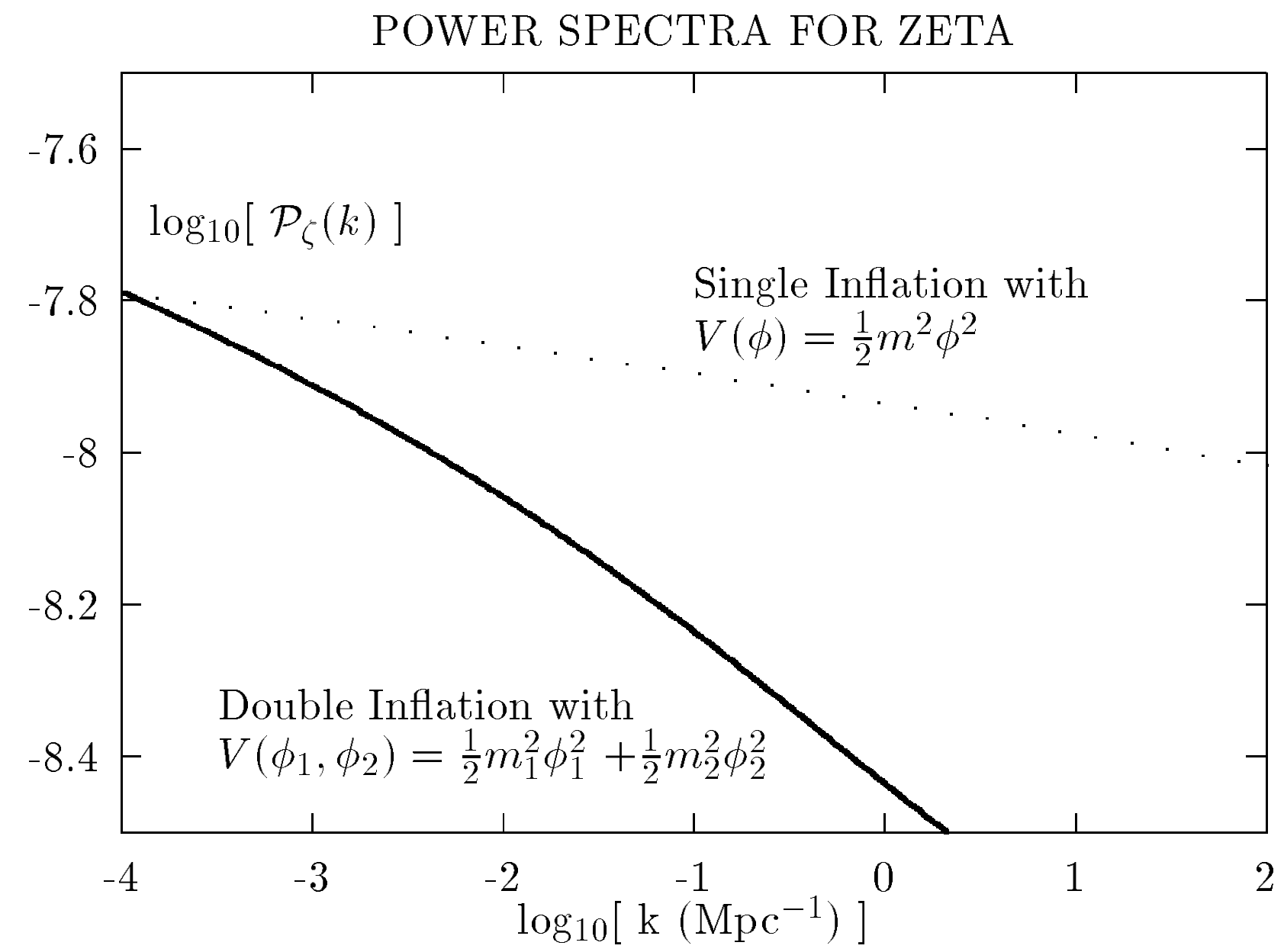
POWER SPECTRA FOR ZETA
$\log_{10}[\ \mathcal{P}_\zeta(k)\ ]$
Single Inflation with
$V(\phi) = \frac{1}{2}m^2\phi^2$
Double Inflation with
$V(\phi_1, \phi_2) = \frac{1}{2}m_1^2\phi_1^2 + \frac{1}{2}m_2^2\phi_2^2$
-7.6
-7.8
-8
-8.2
-8.4
-4
-3
-2
-1
0
1
2
$\log_{10}[\ \mathrm{k}\ (\mathrm{Mpc}^{-1})\ ]$

**Fig.(3a)**

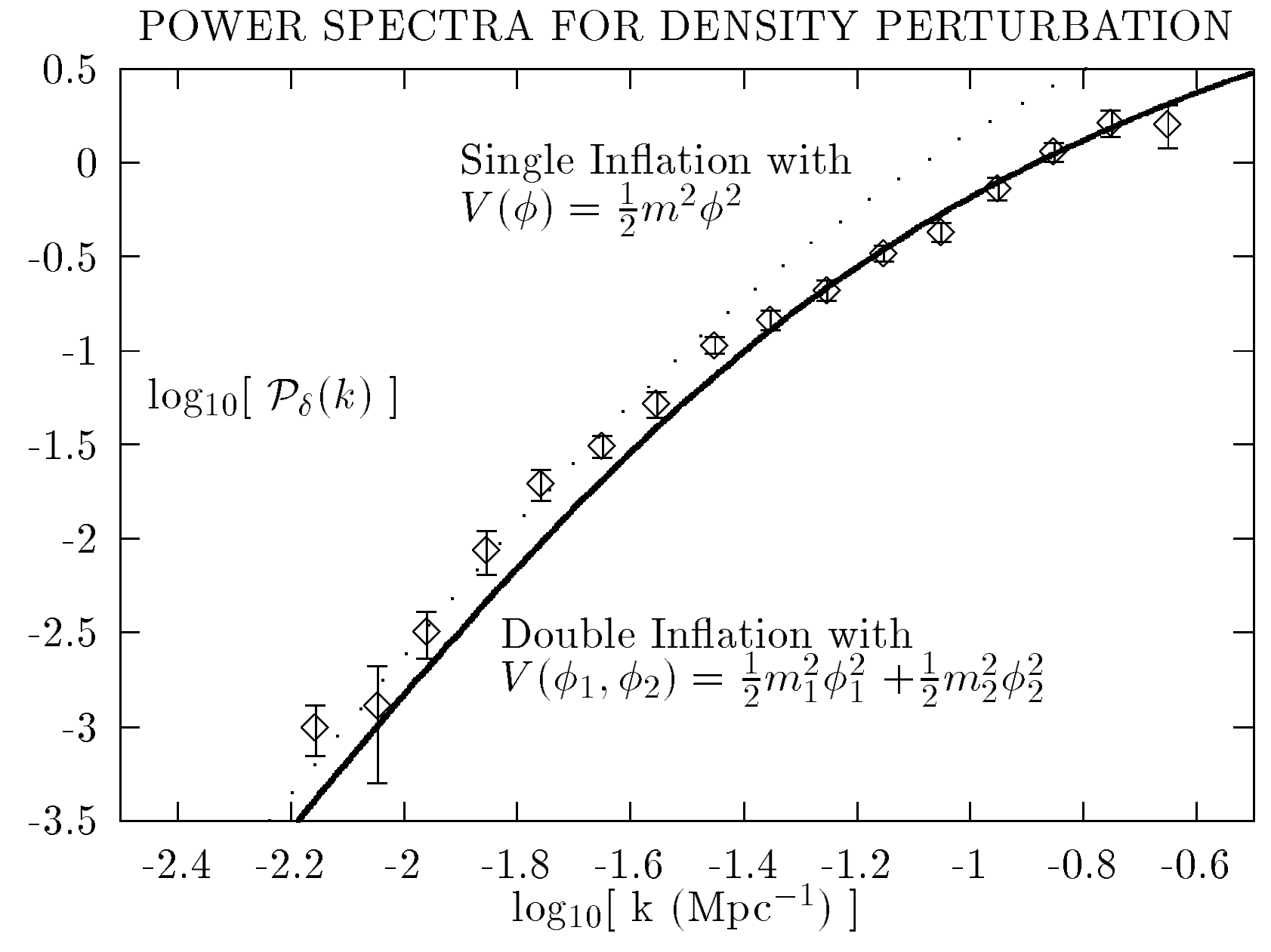
POWER SPECTRA FOR DENSITY PERTURBATION
Single Inflation with
$V(\phi) = \frac{1}{2}m^2\phi^2$
$\log_{10}[\ \mathcal{P}_\delta(k)\ ]$
Double Inflation with
$V(\phi_1, \phi_2) = \frac{1}{2}m_1^2\phi_1^2 + \frac{1}{2}m_2^2\phi_2^2$
0.5
0
-0.5
-1
-1.5
-2
-2.5
-3
-3.5
-2.4
-2.2
-2
-1.8
-1.6
-1.4
-1.2
-1
-0.8
-0.6
$\log_{10}[\ \mathrm{k}\ (\mathrm{Mpc}^{-1})\ ]$

**Fig.(3b)**